\documentclass[aps,prl,groupedaddress,a4paper,preprint,endfloats]{revtex4}
\usepackage{bm,graphicx,graphics,amsmath,amssymb,bm,epsfig,color}
\usepackage{euscript,tabularx}

\begin{document}
%\bibliographystyle{apsrev}
%\draft
\preprint{\begin{minipage}{12cm} Manuscript Number: LS9243 \hrulefill
    (submitted to Phys. Rev. Lett.)
   \end{minipage}}

%cond-mat/0405301
 
\title{Magnetization reduction induced by nonlinear effects}

\author{G. de Loubens}
%\affiliation{Service de Physique de l'{\'E}tat Condens{\'e}, CEA Orme des
%  Merisiers, F-91191 Gif-Sur-Yvette}
\author{V. V. Naletov}
%\affiliation{Service de Physique de l'{\'E}tat Condens{\'e}, CEA Orme des
%  Merisiers, F-91191 Gif-Sur-Yvette}
\altaffiliation[Also at ]{Physics Department, Kazan State University, Kazan 420008 Russia}
\author{O. Klein}
%\thanks{Corresponding author} 
%\email{oklein@cea.fr}
\affiliation{Service de Physique de l'{\'E}tat Condens{\'e}, CEA Orme des
  Merisiers, F-91191 Gif-Sur-Yvette}

\date{\today}

\begin{abstract}
  This letter reports the first detailed measurement of $M_z$, the
  component parallel to the effective field direction, when ferromagnets
  are excited by microwave fields at high power levels.  It is found that
  $M_z$ drops dramatically at the saturation of the main resonance.
  Simultaneous measurements of $M_z$ and absorption power show that this
  drop corresponds to a diminution of the spin-lattice relaxation rate.
  These changes are interpreted as reflecting the properties of
  longitudinal spinwaves excited above Suhl's instability. Similar behavior
  should be expected in spinwave emission by currents.
\end{abstract}


\maketitle

The out of equilibrium dynamics of ferromagnets is receiving much attention
owing to the potential application to spin electronic devices. Interest for
transport properties usually lies in the response of the longitudinal
magnetization $M_z$ (component parallel to the effective field direction)
to a large excitation amplitude. But the high power dynamics cannot be
understood in terms of a linear susceptibility. Nonlinear (NL) contributions
contained in the torque term of the gyroscopic equation become important as
soon as the precession angle exceeds a couple of degrees. Although these
effects were first discovered in the ferromagnetic resonance (FMR) response
of insulators \cite{damon:53}, they are in fact generic to all
ferromagnets. The same physics also applies to metallic samples as shown
recently by An \textit{et al.}  \cite{an:04}.  While the consequences of
these nonlinearities on the microwave susceptibility have been thoroughly
investigated \cite{suhl:57}, their effects on $M_z$ have never been
established.

This letter reports the first measurement of $M_z$ at the resonance
saturation.  We will present our data obtained at room temperature on an
yttrium iron garnet (YIG) sample, shaped into a disk of diameter $D=160 \mu$m
and thickness $4.75 \mu$m and uniformly magnetized by a static magnetic
induction $B_\text{ext}$ applied along the normal axis $z$ of the disk
\cite{naletov:03}.  It is found that $M_z$ drops dramatically at the
saturation of the main resonance. It results from a rapid growth of
phase-coherent spinwaves (SW) propagating along the magnetization direction
\cite{suhl:57}. We use these experimental results to discuss the
consequences of the NL effects on the problematic of spin injection.
Their inclusion inside the microscopic spin-torque model
\cite{slonczewski:96,berger:96,waintal:00} should provide a mechanism to
reverse the magnetization through a reduction of $M_z$
\cite{silva:02,safonov:04}.
 
\begin{figure}
\includegraphics[width=8cm]{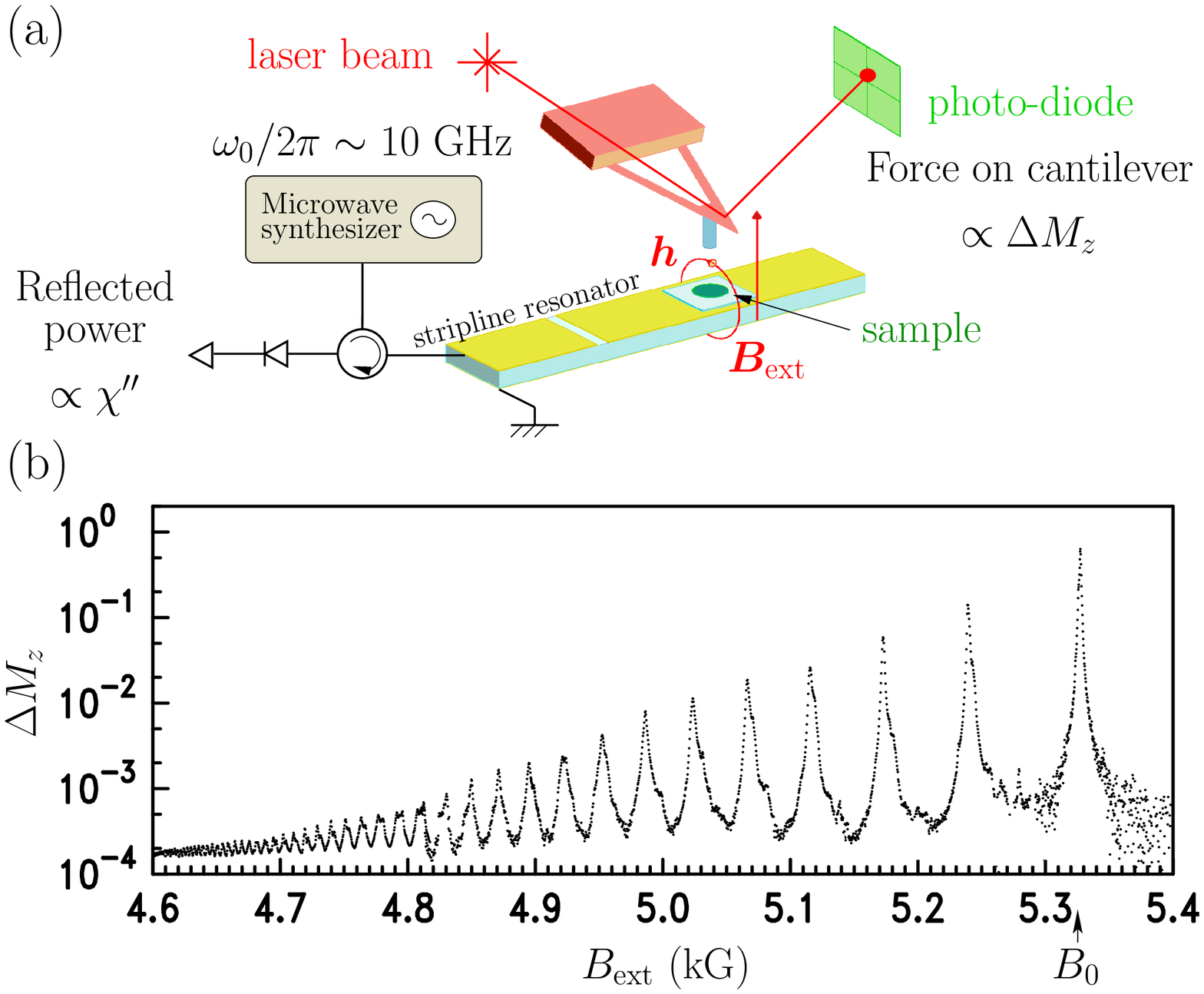}
\caption{
  (a) Schematic of the experimental setup. A Magnetic Resonance Force
  Microscope is used to measure $\Delta M_z$ while a standard microwave setup
  measures $\chi''$. (b) Linear FMR spectrum of the micron-size YIG disk
  detected mechanically. The main resonance occurs at $B_0$.}
\label{fig1}
\end{figure}

We exploit the exquisite sensitivity of Magnetic Resonance Force Microscopy
(MRFM) to follow the changes in $M_z$ with a resolution better than 1ppm.  A
schematic of the setup is shown in Fig.\ref{fig1}(a).  A cylindrical
permanent magnet is glued at the free end of a clamped cantilever and then
aligned with the axis of the YIG disk. The distance between the sample and
the probe is fixed at $100\mu$m so that their coupling is in the weak
interaction regime \cite{charbois:apl}. The inhomogeneous dipolar field
generated by the sample creates a point load on the tip and thus an elastic
deformation of the cantilever.  Since the flexural modes of the cantilever
are well below the Larmor frequency, the mechanical probe is insensitive to
the precession of the transverse magnetization and it only couples to
the longitudinal component $M_z$.

The measured quantity is $\Delta M_z = M_s -M_z$ when the microwave field is
turned on. $M_s$ is the saturation magnetization at the lattice
temperature. The microwave is generated by a synthesizer and fed into an
impedance matched stripline resonator tuned at $\omega_0/2 \pi = 10.47$GHz. We
call $h$ the circularly polarized amplitude of the microwave driving field
at the sample position. If $h$ is much lower than Suhl's threshold (see
below) then the response is proportional to the excitation power.
Fig.\ref{fig1}(b) shows the FMR spectrum of our disk in the linear regime.
A multiplicity of $\Delta M_z$ maxima are detected during a sweep of
$B_\text{ext}$. The spectrum illustrates the quantized energy levels of
magnetostatic waves (near-zero wavevector) confined by the sample diameter.

Simultaneously, we measure the imaginary part of the transverse
susceptibility $\chi''$ through a standard setup. The power reflected off the
half-wavelength resonator is detected by a microwave crystal diode whose
signal is proportional to the absorbed power $P_\text{abs} = \omega_0 \chi''h^2$.

\begin{figure}
\includegraphics[width=9cm]{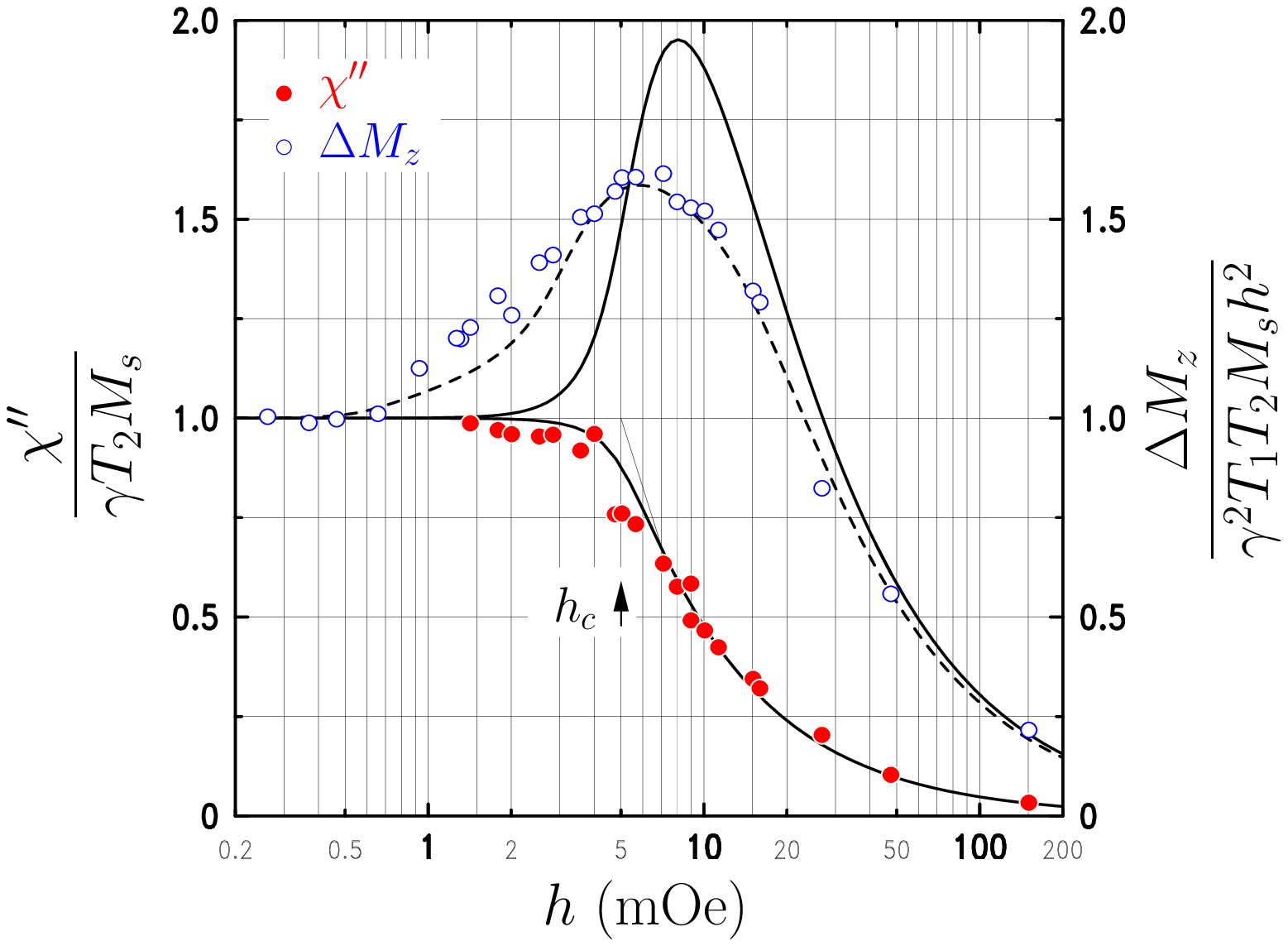}
\caption{
  Microwave field strength dependence of the transverse (closed circles)
  and longitudinal (open circles) components of the magnetization at $B_0$.
  The solid lines are the analytical predictions (see text). The dashed
  line is the behavior corrected by finite size effects.}
\label{fig2}
\end{figure}

Although the discussion below can be extended to any of the magnetostatic
modes above, we concentrate on the high power behavior of the main
resonance at $B_0=5324.5$G (see Fig.\ref{fig1}(b)). This mode has a
transverse wavevector $k_0 \approx \pi/D$ and is called hereof the uniform
precession because it has no precessional nodes. We plot in Fig.\ref{fig2}
the microwave field dependence of $\chi''$ (closed circles) and $\Delta M_z/h^2$
(open circles) evaluated at resonance. Both quantities are normalized by
their low power values. The $\chi''$ data shows the well known premature
saturation behavior above Suhl's threshold, $h_c = 5$mOe.  Surprisingly,
the power dependence of $\Delta M_z/h^2$ exhibits a peak at the threshold.
Evidence of a behavior where $\Delta M_z/h^2$ increases with power has never
been established before in any magnetic resonance experiment. 
 
\begin{figure}
  \includegraphics[width=8cm]{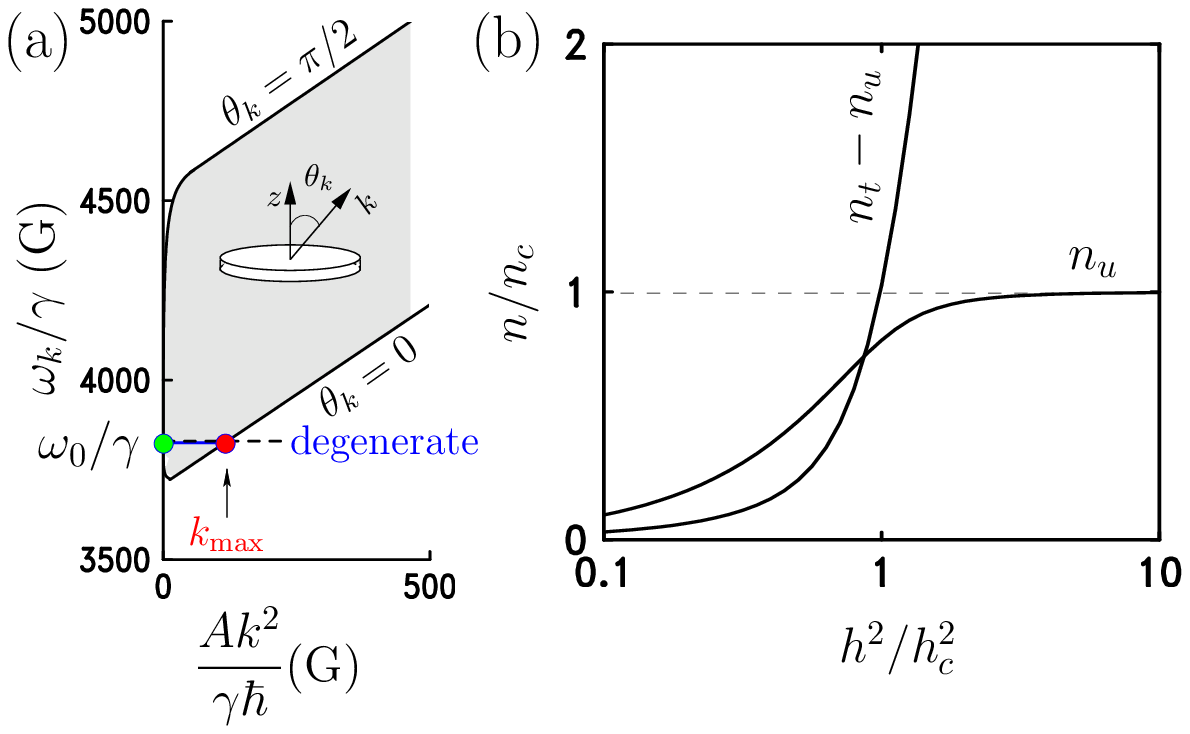}
\caption{
  (a) SW dispersion: the shaded area is the magnon manifold
  explained in \cite{sparks} ($A$ is the exchange constant). (b) Power
  dependence of $n_u$ and $n_t-n_u$ for our sample.}
\label{fig3}
\end{figure}

Obviously the correlation between the longitudinal and transverse
quantities cannot be explained by a macrospin picture, which implies that
$\Delta M_z = \chi''^2 h^2/(2 M_s)$ in this power range.  The difference clearly
demonstrates that the $\chi''$ measurements only reveal a partial picture of
the dynamics inside the sample. Other SW with identical energies ($\omega_k =
\omega_0$) but different wavevectors ($k \gg k_0$) can be excited in the sample.
Fig.\ref{fig3}(a) is the magnon-manifold for our finite aspect ratio disk
\cite{sparks}. The shaded area expresses the possible balance between
demagnetizing and exchange energies. Volume demagnetizing energy of SW
mode can be changed by varying the angle $\theta_k$ between $\bm{k}$, the
propagation wavevector and $\bm{M}_s$, the magnetization vector.  Short
wavelength SW do not couple to $\chi''$ because the transverse projection
averages to zero.  However, each magnon excitation diminishes the
longitudinal component by $\gamma \hbar$. In other words, while the transverse data
measures the number of uniform magnons $n_u=\frac{1}{2} \frac{\chi''^2h^2}{M_s
  \gamma \hbar}$, the longitudinal data is proportional to the total number of
magnons $n_t= \frac{\Delta M_z}{\gamma \hbar}$. Here we only count the magnons created by
the microwave excitation. Thermal magnons' occupation numbers are accounted
for in $M_s$ and remain constant in all the measurements below.  Suhl
showed \cite{suhl:57} that the NL coupling destroy the independence between
degenerate modes.  He found that, above a critical power ($h_c^2$), $n_u$
saturates at $n_c$ as shown in Fig.\ref{fig3}(b). We want to establish that
this saturation is associated with a surge of degenerate magnons $n_t-n_u$.

\begin{figure}
  \includegraphics[width=8cm]{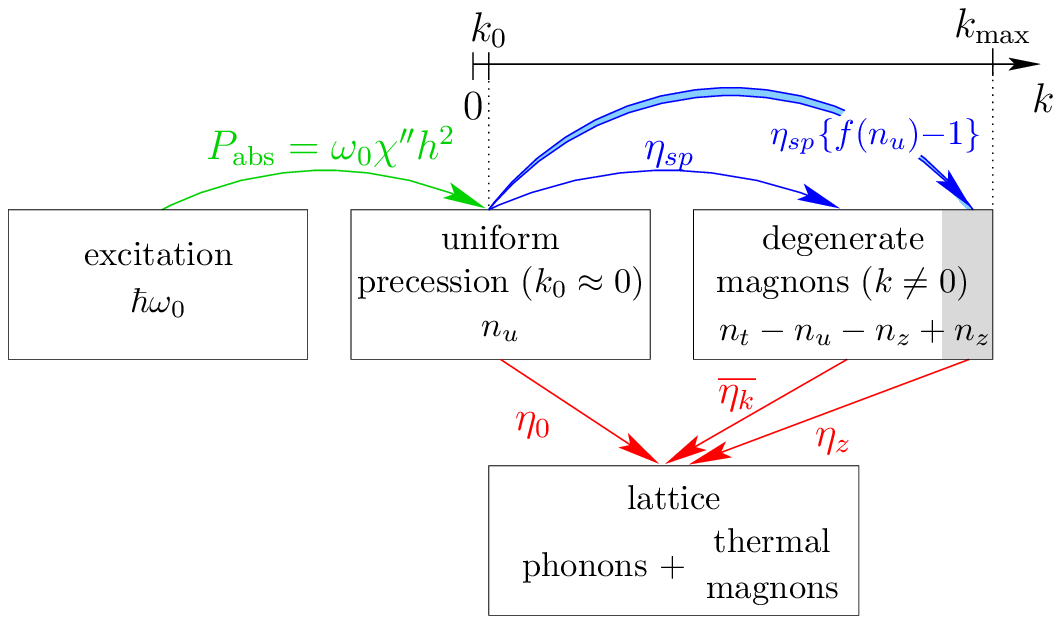}
\caption{
  Block diagram of the energy transfers between the various degrees of
  freedom of the system. Transfers between the upper reservoirs conserve
  the total energy. The linear regime corresponds to $f(n_u) =1$. The
  shaded area and doubled line are the alterations by NL effects.}
\label{fig4}
\end{figure}

Before continuing further the discussion, we need to look at the different
relaxation channels in our sample. These damping forces are necessary to
stabilize the SW excited by the nonlinearities \cite{suhl:57}.
Fig.\ref{fig4} establishes a schematic of the coupling between the various
degrees of freedom in the system.  At $\omega_0$ and $B_0$, the uniform
microwave field preferentially couples to the uniform mode, the longest
wavelength mode available inside the degenerate band.  Its energy
relaxation rate directly to the lattice will be written $\eta_0$ (our
definition is twice as large as the amplitude relaxation rate used in
Suhl's paper \cite{suhl:57}). The linewidth of the peaks in
Fig.\ref{fig1}(b), however, is the sum $(\eta_0 + \eta_{sp})/ \gamma$, where $\eta_{sp}$
is the decay constant of the uniform precession to degenerate SW due to
scattering on the sample inhomogeneities.  We define $n_k$ as the number of
degenerate magnons having a wavevector $k$ and $\overline{\eta_{k}}$ as their
average decay rate to the thermodynamic equilibrium.  Translated into
Bloch's notation, the linear part of the diagram in Fig.\ref{fig4}
corresponds \cite{fletcher:60} to $T_2= 2/(\eta_0 + \eta_{sp})$ and $T_1 = T_2/2
\left ( 1 + \eta_{sp} / \overline{\eta_k} \right )$, respectively the transverse
and longitudinal relaxation times. These relaxation times, in the limit of
infinitesimal excitations, have been completely characterized in a previous
paper \cite{klein:03}. Their values are summarized in Table \ref{tab1}.

\begin{table}
\begin{tabular}{c | c c c}
\hline \hline 
% & \multicolumn{3}{c}{energy relaxation} \\
spin-spin &\  uniform \ &\   $k$th\ &\  longitudinal  \\
process  & precession    & magnons        & magnons    \\
\hline\hline\\
$\eta_{sp}/ \gamma$ & $\eta_0/ \gamma$   & $\overline{\eta_k}/ \gamma$  &  $\eta_z/ \gamma$  \\
\hline \\
0.2 & 1.07  & 0.65 & 0.15 \\
\hline\hline
\end{tabular}
\caption{Relaxation rates of degenerate magnons (in G).}
\label{tab1}
\end{table}

The NL terms in the gyroscopic equation \cite{suhl:57} couple
coherently degenerate SW of equal and opposite wavevectors $(+k,-k)$.  The
coupling term with the uniform motion, $\xi_{k} n_u$, is maximum
($\xi_{k}|_\text{max}/ \gamma = 2 \pi M_s \approx 900$G) for longitudinal SW ($\theta_k =0$ or
$\bm{k} \| \bm{M}_s$). We use Suhl's notation for $\xi_{k}$ so that the reader
can refer to \cite{suhl:57} for a complete treatment. In our disk, these SW
have a wavevector $k_\text{max} \approx 6.3 \times 10^4 \text{cm}^{-1}$ (see
Fig.\ref{fig3}(a)). A critical threshold is reached when $\xi_{k} n_u$
becomes comparable to $\eta_{z}$, the relaxation rate to the lattice of the
longitudinal SW. Two quanta of the uniform precession then break down in a
parametric magnon pair propagating along $\bm{M}_s$. The instability
corresponds to a spatial distortion of the instantaneous axis of precession
which diminishes the transverse demagnetizing energy. The critical number
of uniform magnons is $n_c =\frac{1}{2} \frac{ \gamma M_s T_2^2 h_c^2}{\hbar} $ with
$h_c^2 =\frac{1}{\gamma^2 T_2^2} \frac{\eta_{z}}{2\xi_k}$, the saturation power. We
propose to view this as an alteration of the energy flow between the modes
in Fig.\ref{fig4}:
\begin{subequations} 
%\left \{
\begin{eqnarray} 
 \frac{\partial }{\partial t}n_u   =  \frac{ P_\text{abs} }{\hbar \omega_0}
  - \left \{\eta_0 + \eta_{sp}  f(n_u) \right \} n_u \label{nu} \\
 \frac{\partial  }{\partial t}(n_t - n_u -n_z) =   \eta_{sp} n_u  - \overline{\eta_k}  \left(
  n_t - n_u - n_z \right ) \\
 \frac{\partial }{\partial t} n_z  =  \eta_{sp}  \left \{ f(n_u) -1 \right \} n_u - {\eta_z} n_z,
  \end{eqnarray} \label{nt}
%\right .
\end{subequations}
where $f(n_u) = 1/\sqrt{1 - n_u^2/n_c^2}$ \cite{suhl:59} and $n_z$ is the
number of parametric magnons.

In the stationary regime, Eq.\ref{nu} leads to an implicit equation for the
susceptibility first derived by Suhl \cite{suhl:59}:
\begin{equation}
\widetilde{\chi ''} = \frac{\eta_0 + \eta_{sp}}{\eta_0 + \eta_{sp}/ \sqrt{1 - 
      \widetilde{\chi ''}^4  \left (h/h_c \right )^4}},  \label{susc}
\end{equation}
where $\widetilde{\chi ''} = \chi ''/(\gamma T_2 M_s)$ is the susceptibility
normalized by its low power value.  We try to compare quantitatively our
data to the profile predicted by Eq.\ref{susc}. We use the measured value
of the threshold, $h_c = 5$mOe, in the formula. It corresponds to $\eta_{z} /
\gamma =0.15$G, a value which is consistent with previous measurements in
spheres and thin plates \cite{patton:79}, where it was found that the decay
rate becomes smaller as $\theta_k$ diminishes \cite{patton:79,fletcher:60}.  We
have displayed in Fig.\ref{fig2} the predicted behavior, with no fitting
parameters.  The model properly predicts the shape and the smearing out of
the singularity at $h_c$.

Solving the set of Eq.\ref{nt} in the steady-state gives a new analytical
relationship between ${\chi''}$ and $\Delta M_z/h^2$
\begin{equation}
\frac{ \widetilde{\Delta M_z}}{h^2} = \frac{1}{T_1 \eta_{z} } \ 
 \left \{ \widetilde{\chi ''} - \widetilde{\chi ''}^2 \right \} +  \widetilde{\chi ''}^2,
\label{equilibrium} 
\end{equation}
where $\widetilde{\Delta M_z} = \Delta M_z/(\gamma^2 T_1 T_2 M_s)$ is normalized by the
slope of $\Delta M_z/h^2$ at low power.  The above equality is equivalent to an
energy balance between the absorbed power and the dissipated power in the
sample $P_\text{abs}=P_\text{diss} = \hbar {\omega_0} \sum_{\{k\}} \eta_k n_k$
\cite{schlomann:59,fletcher:60}. The solid line in Fig.\ref{fig2} is the
bell-shape behavior inferred from Eq.\ref{equilibrium} when $T_1 \eta_{z} =
0.15$.  Experimentally we find that the raise of $\Delta M_z/h^2$ begins at
powers lower than $h_c$. We have omitted in our model the finite size
effects in order to keep the discussion simple.  But the so-called uniform
mode has a spatially dependent precession amplitude.  $\Delta M_z$ at the center
of the disk is about three times larger than its spatial average
\cite{naletov:03}.  Changes in the profile are expected when different
locations of the sample hit the saturation threshold.  Spatial dependences
are further magnified by the MRFM which provides a local measurement (here
the mechanical probe is above the disk center). While the raise of $\Delta
M_z/h^2$ coincides with the saturation of the central part of the disk, the
spatially averaged transverse susceptibility $\chi''$ drops at higher power,
approximately when the periphery saturates (zone with the highest spatial
weight).  The dashed line in Fig.\ref{fig2} is the predicted behavior when
Eq.\ref{equilibrium} is weighted by the spatial profile of $\Delta M_z$. Such
good agreement with the data suggests that the shift of the peak towards
$h<h_c$ is due to these finite size effects.

\begin{figure}
\includegraphics[width=7cm]{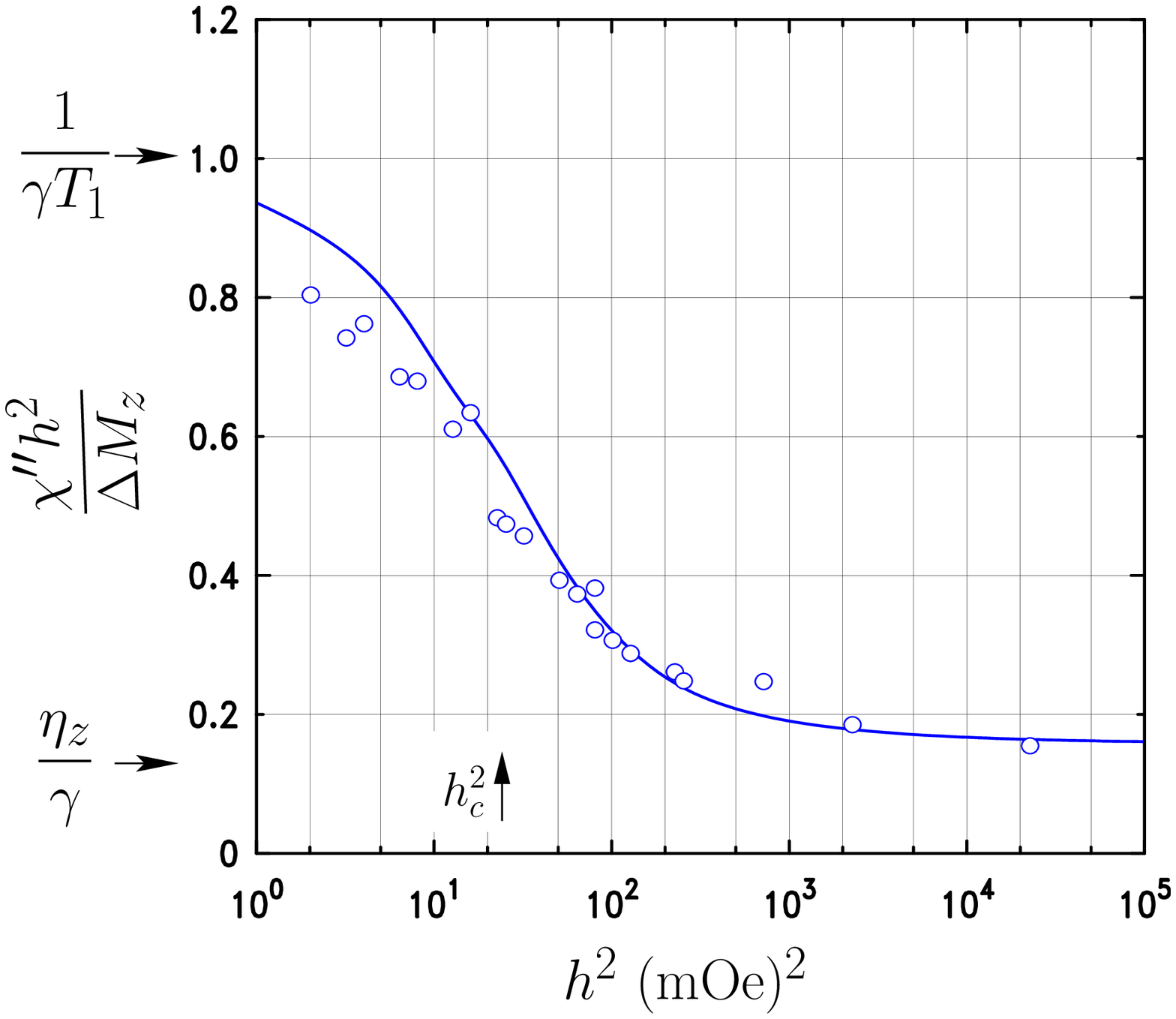}
\caption{ 
  Power dependence of the ratio of $\chi''$ over $\Delta M_z/h^2$. It measures the
  energy decay rate of the spinwave system. The solid line is the
  analytical prediction.}
\label{fig5}
\end{figure}

To gain further insight, we plot in Fig.\ref{fig5} the ratio between the
transverse and longitudinal quantities. This shows the power dependence of
the spin-lattice relaxation rate inside the spin system. In magnetic
resonance, the power absorbed during the spin-lattice relaxation time is
equal to the energy stored in the sample, $\frac{\omega_0}{\gamma} {\Delta M_z}$.  Thus
the quantity on the ordinate axis is $ \frac{1}{\omega_0} \frac{P_\text{abs}}{\Delta
  M_z} = \frac{1}{\gamma} \sum_{\{k\}} \eta_k \frac{n_k}{n_t}$.  We find that the
observed correlation between $\chi''$ and $\Delta M_z/h^2$ corresponds to a
monotonic diminution of the damping with increasing power. This ratio
starts to decrease below $h_c$ and the drop extends to the $h>h_c$ region.
In our model, the relaxation rate is simply equal to $1/T_1$ when $f(n_u)
=1$ (linear regime) and then it reduces to $\eta_z$ when $f(n_u)\gg 1$.  The two
limits correlate respectively to $n_t \approx n_u$ and $n_t \approx n_z$. The solid
line in Fig.\ref{fig5} is the drop inferred from the analytical expression
showing how the $\eta_z$ value is approached asymptotically. We find that the
agreement with the data extends well above $h_c$ and the asymptotic value
is in agreement with the $\eta_z$ inferred from the threshold value, $h_c$.
We recall that our model only considers the NL coupling between the uniform
and the longitudinal magnons. A proper analysis, however, should also
account for the NL coupling between all the degenerate modes, which become
important at much higher powers. They correspond to a different
redistribution of the degenerate magnons' occupation number in the flow
diagram of Fig.\ref{fig4}.

In conclusion, this letter studies the changes of properties of excited SW
due to NL effects. If the excitation mechanism favors a particular mode,
then a 4-magnon process starts at high power levels to redistribute the
magnons' occupation number between degenerate modes. During the
redistribution, the coupling to the excitation usually decreases: less
energy is absorbed for the same excitation energy.  The efficiency,
however, might increase if the newly filled modes (generally the
longitudinal ones \cite{sparks}) have lower energy relaxation rates: more
SW are emitted for the same absorbed energy.  We believe that these
comments are also pertinent to SW emitted by a current for the following
reasons. First, the above statements can be generalized to other selection
rules, \textit{i.e.}  the favored mode might not necessarily be the longest
wavelength one.  Second, the precession amplitudes achieved by currents are
much larger than the ones achieved by microwaves \cite{kiselev:03} (though
conductors have larger values of $\eta_0$). Third, the eigen-modes are at
least doubly degenerate if one sample dimension is larger than 100nm
(indeed the width of the magnon manifold is of the order of $2\pi/
k_\text{max} \lesssim 100$nm for metals). Fourth, the energy relaxation rate of
degenerate SW generally decreases as $\bm{k}$ aligns with the $\bm{M}_s$
direction (cf.  measurements in Permalloy films \cite{an:04}).  These
arguments indicate that parametric SW should also get excited by spin
injection.  Excitations of short wavelength SW should provide a mechanism
to reverse the magnetization through a reduction of $M_z$
\cite{safonov:04}. Finally, we emphasize that the NL transfers conserve the
phase-coherency. This point is important regarding to the spin torque model
which relies on the concept of coherent emission of SW \cite{berger:96}.

We are greatly indebted to J. Ben Youssef, V.  Charbois, M. Viret, X.
Waintal, C.  Fermon and O.  Acher for their help and support. This research
was partially supported by the E.U. project M$^2$EMS (IST-2001-34594) and
the Action Concert{\'e}e Nanoscience NN085.

%\vspace{-0.7cm}
\bibliography{/home/klein/latex/bib/fmr,/home/klein/latex/bib/vcthesis,/home/klein/latex/bib/mrfm,/home/klein/latex/fmrfm/suhl/notes,/home/klein/latex/bib/spin}

%\bibinfo{journal}{{to be published in J. Appl. Phys. (preprint available by email: patton@lamar.colostate.edu)}}

\end{document}